\documentclass[11pt,twoside]{article}

\usepackage{baaa2013}

\usepackage{graphicx}
\usepackage{subfigure}
\usepackage{amssymb}
\usepackage[spanish,activeacute,english]{babel}
\usepackage[latin1]{inputenc}
\usepackage{verbatim}
\usepackage{amsmath}
\usepackage{amsfonts}
\usepackage{amssymb}
\usepackage[colorlinks=true,dvips]{hyperref}
\usepackage{natbib}

\usepackage{rotating}

\begin{document}
\myselectspanish
\vskip 1.0cm
\markboth{Catelan et al.}%
{The VVV Survey: An Update}

\pagestyle{myheadings}
\vspace*{0.5cm}


\Categ{3}


\Tema{2}

\vskip 0.3cm
\title{Stellar Variability in the VVV Survey: An Update}

 
\author{M. Catelan$^{1,2}$, I. D\'ek\'any$^{2,1}$, M. Hempel$^{1,2}$ \& D. Minniti$^{1,2}$}

\affil{%
  (1) Pontificia Universidad Cat\'olica de Chile, Facultad de F\'{i}sica, Instituto de Astrof\'{i}sica, Santiago, Chile\\ 
  (2) Millennium Institute of Astrophysics, Santiago, Chile\\ 
}

\vskip 0.5cm

\begin{abstract}
The {\em Vista Variables in the V\'{i}a L\'actea} (VVV) ESO Public Survey consists in a near-infrared time-series survey of the Galactic bulge and inner disk, covering 562~square degrees of the sky, over a total timespan of more than 5~years. In this paper, we provide an updated account of the current status of the survey, especially in the context of stellar variability studies. In this sense, we give a first description of our efforts towards the construction of the {\em VVV Variable Star Catalog} (VVV-VSC).  
\end{abstract}

\vskip 0.5cm

\begin{resumen}
El Relevamiento P\'ublico ESO titulado {\em Variables Vista en la V\'{i}a L\'actea} (VVV) consiste en un relevamiento con resoluci\'on temporal del bulbo y disco interno Gal\'acticos en el infrarrojo cercano, cubriendo 562~grados cuadrados del cielo, en un intervalo de tiempo de m\'as de 5~a\~nos. En este art\'{i}culo, entregamos una descripci\'on actualizada del estado actual del relevamiento, especialmente en el \'ambito de los estudios de variabilidad estelar. En ese sentido, describimos, por primera vez, nuestros esfuerzos para la construcci\'on del {\em Cat\'alogo de Estrellas Variables} del VVV (VVV-VSC).   
\end{resumen}

\section{Overview}\label{sec:intro}

\noindent The Vista Variables in the V\'{i}a L\'actea (VVV) ESO Public Survey \citep{dmea10,mcea11,rsea12,mhea14} 
is a time-series, near-infrared (IR) survey of the Galactic bulge and an adjacent portion of the inner 
disk, covering 562~square degrees of the sky. The survey, which is based on images taken with the VIRCAM camera mounted on the VISTA telescope, located in Cerro Paranal, northern Chile, has provided
multi-color photometry in 5 broadband filters ($Z$, $Y$, $J$, $H$, and $K_s$), but its main goal 
is to provide a homogeneous database for variability studies of the observed regions, particularly in the $K_s$ band. 

An extensive overview of recent work on stellar variability that has been carried out in the framework of the VVV Survey has been presented by \citet{mcea13}. Here we provide a brief update on the status of the survey, particularly in regard to our ongoing work on stellar variability.

\section{Status of the Survey}\label{sec:status}

\begin{sidewaysfigure}
    \centering
    \includegraphics[width=\textheight]{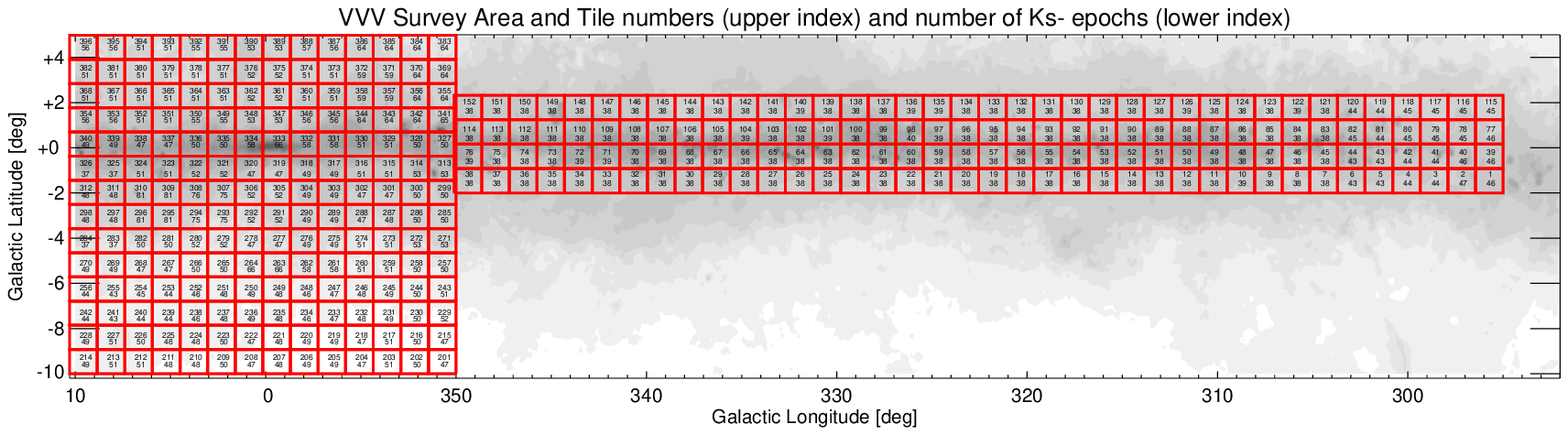}
\caption{VVV Survey area, superimposed on a gray-scale image that represents the Galactic dust extinction maps by \citet{schleg98}. Each of the 348 tiles of the survey is represented by a red rectangle, and is identified by its ID number ({\em upper label} within each tile). Each such tile, which covers a $\approx$~1.5 square degree field, has already been observed at least once in each of the $Z$, $Y$, $J$, $H$, and $K_s$ filters. The actual number of epochs in the $K_s$-band as of this writing (May 16, 2014) is shown underneath each tile's ID number.}
\label{fig:area}
\end{sidewaysfigure}

\begin{figure*}[!t]
\centering \includegraphics[width=0.86\textwidth]{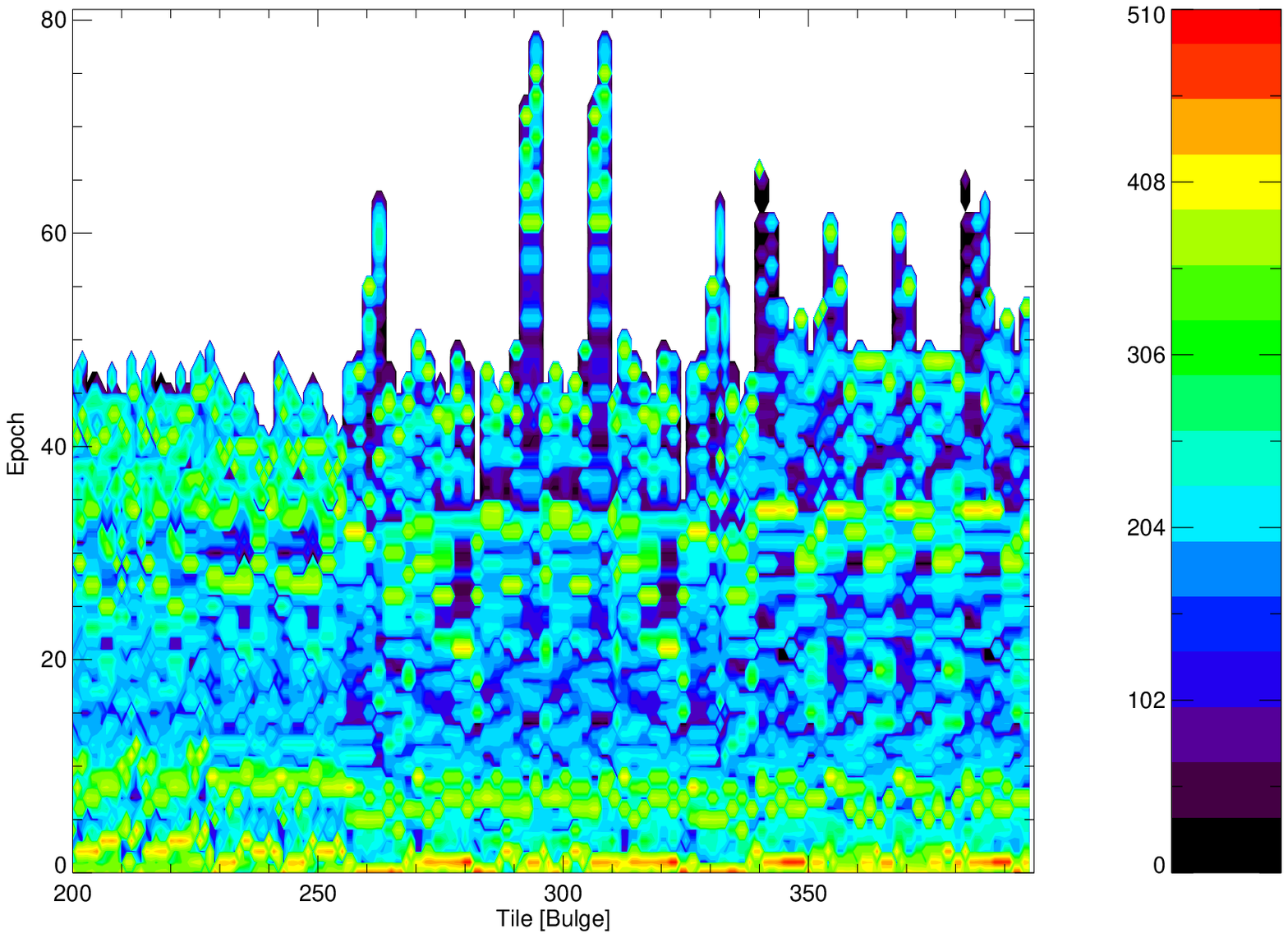}
\centering \includegraphics[width=0.86\textwidth]{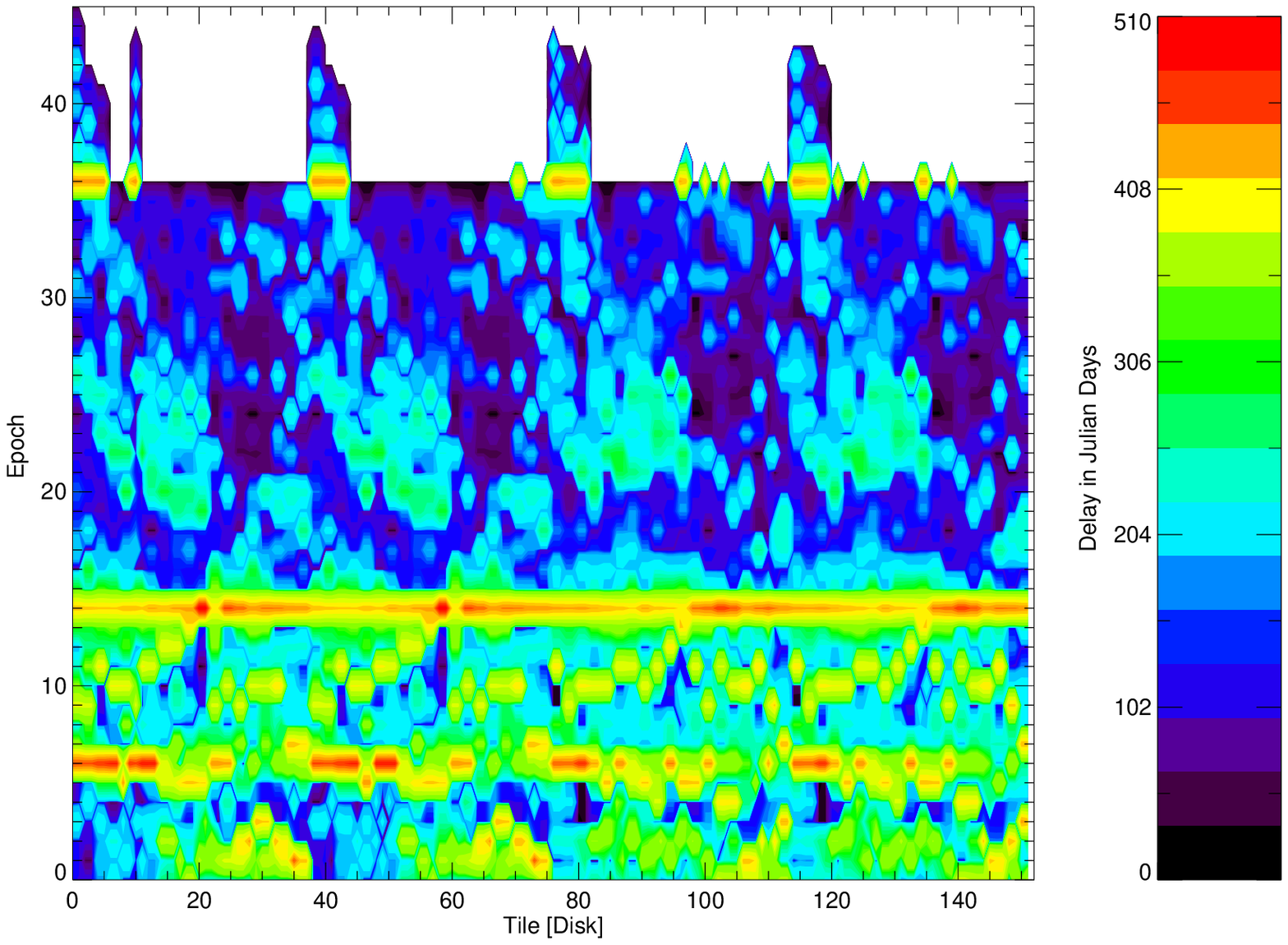}
\vskip0pt
\caption{Time delay (in Julian Days) between two consecutive epochs of $K_s$-band
  observations for the bulge ({\em top}) and disk ({\em bottom}) tiles, based on all VVV time-series data obtained as of this writing (May 16, 2014). The 8 tiles with the largest number of observed epochs (ID numbers: 293, 294, 295, 296, 307, 308, 309, 310; see also Fig.~\ref{fig:area}) are bulge tiles situated in Baade's Window, with a much reduced  dust extinction and for which independent observations are available in the optical bands from the OGLE survey \citep{auea92}.}
\label{fig:epochs-delays}
\end{figure*}

\noindent Figure~\ref{fig:area} shows the complete area covered by VVV, with the different ``tiles'' (which constitute individually observed fields) indicated as red squares. Each tile covers 1.501~deg$^2$ in the sky, and the survey encompasses 348 tiles in total. For each VVV tile, Figure~\ref{fig:area} also shows its ID number ({\em upper label}) and the number of epochs in $K_s$ that have been obtained, as of this writing ({\em bottom label}). As far as the cadence, Figure~\ref{fig:epochs-delays} shows the delay (i.e., time interval) between two successive epochs in the VVV $K_s$-band data, color-coded according to the scale on the right of each plot. By the end of the survey, the bulge region will contain between 60 and 100 epochs for each tile, whereas the disk tiles will have $\approx 60$~epochs each.  

\begin{figure*}[!t]
\centering \includegraphics[width=\textwidth]{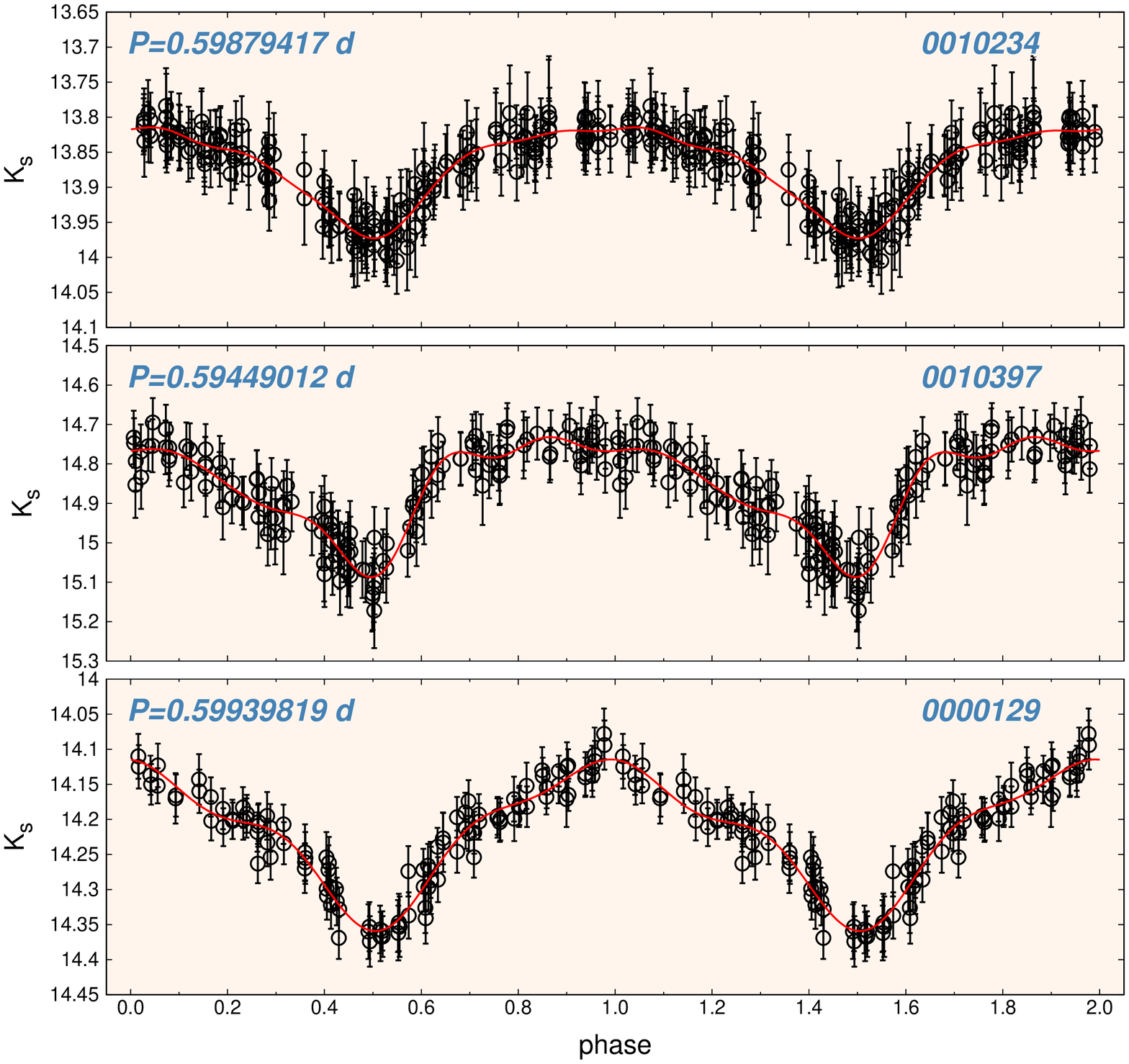}
\vskip0pt
\caption{These panels show the $K_s$-band phase diagrams of 3 bulge ab-type
(i.e., fundamental-mode) RR~Lyrae stars 
known from the OGLE-III survey (\citealt{isea11}; OGLE IDs and periods are 
shown for each star), illustrating the high quality of the data obtained in 
the course of the VVV Survey.  
Note that these three stars have
very similar periods, yet quite different light-curve shapes. The {\em red
lines} show the Fourier fits with optimized numbers of terms. The figure
includes data obtained up to the end of June 2013.
}
\label{fig:3rrl}
\end{figure*}

Figure~\ref{fig:3rrl} shows some examples of VVV light curves for bulge 
RR~Lyrae stars. These stars had been previously identified by the OGLE-III 
survey \citep{isea11}, and so their pulsation periods are accurately known. 
The VVV light curves of known RR~Lyrae with very accurately known periods 
will enable us to systematically investigate, with unprecedented
detail, the near-infrared light-curve properties of these stars: note, 
for instance, how the three RR~Lyrae stars shown in Figure~\ref{fig:3rrl} 
have different light-curve morphologies, in spite of their very similar 
periods. Such detailed studies of light-curve morphology will 
be important for constraining pulsation models \citep[e.g.,][]{gbea00}. 

Using such well-defined $K_s$-band light curves, 
we will also study possible relationships between near-infrared light-curve
parameters and the metallicity (in addition to stellar physical parameters),  
as has been done many times previously 
in the optical for different types of variables, most notably Cepheids and  
RR~Lyrae stars \citep[e.g.,][]{sl81,st82,jk96,kw01,smea07,jnea13}. 
In turn, this might prove very useful for studying the
metallicity distributions/gradients of the stellar populations that
these stars trace~~-- i.e., in the bulge and the thick disk, in the case 
of RR~Lyrae stars \citep[e.g.,][]{al95}.

In like vein, by combining RR~Lyrae light curves from visual databases and 
VVV light curves, we are in a position to simultaneously derive extinction 
and distance {\em to each individual star}. This can be achieved by applying the 
RR~Lyrae period-luminosity calibrations in the visual and near-infrared bands, such 
as provided by \citet{mcea04} in $I$, $J$, $H$, and $K$. A first such application 
has recently been described by \citet{idea13}, where bulge RR~Lyrae stars in 
common between the OGLE and VVV surveys were used, together with the 
\citeauthor{mcea04} period-luminosity relations, to infer an RR~Lyrae-based 
distance to the Galactic center, estimated at $R_0 = 8.33 \pm 0.05 \pm 0.14$~kpc, 
where the indicated error bars correspond to the estimated statistical and systematic  
uncertainties, respectively. This is in remarkable agreement with the dynamical 
value derived by \citet{sgea09}, who have obtained $R_0 = 8.35 \pm 0.35$~kpc
by studying the orbits of stars around the Galaxy's central black hole.

\begin{figure*}[!t]
\centering \includegraphics[width=\textwidth]{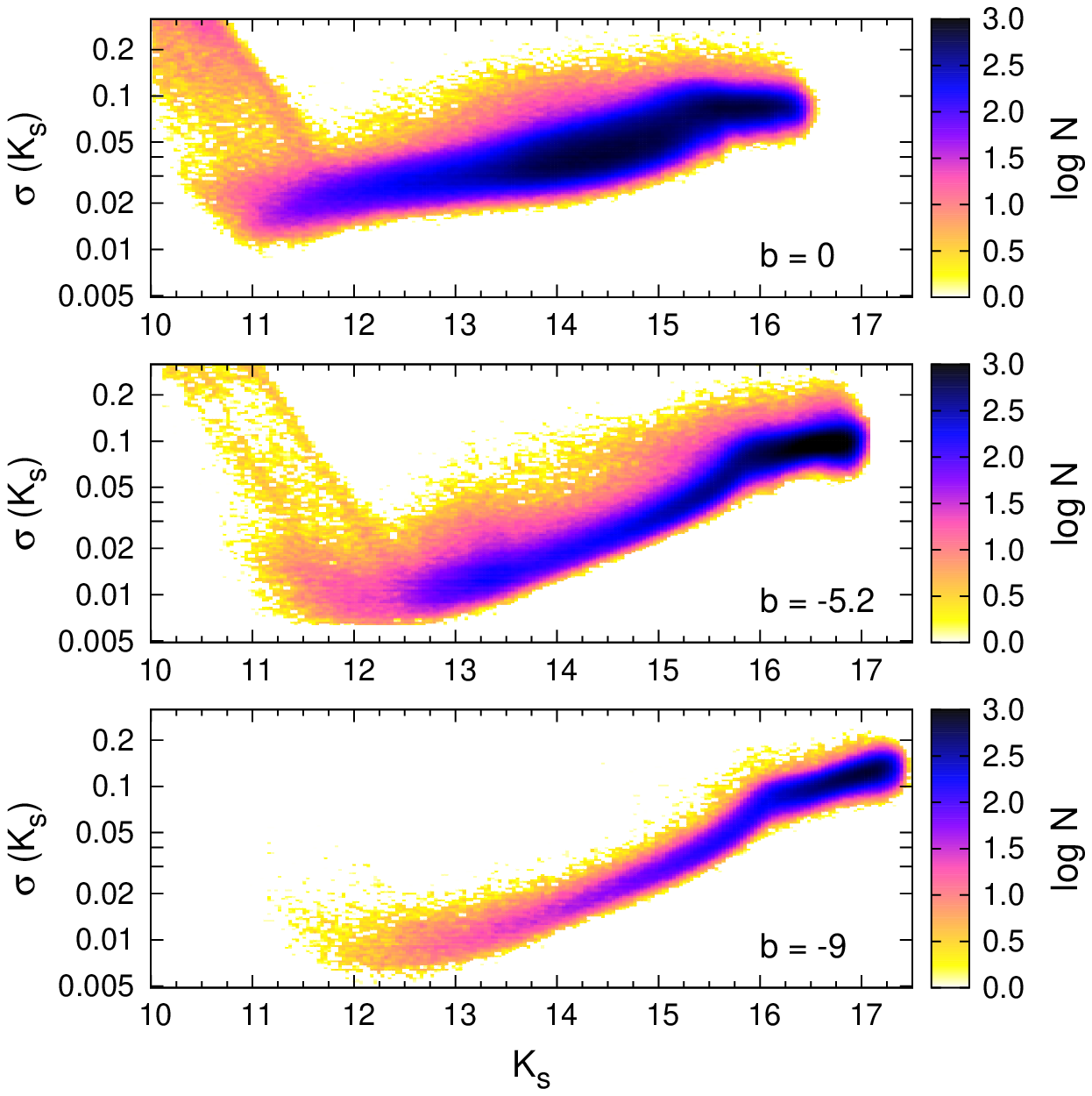}
\vskip0pt
\caption{This figure shows the 
number density distribution of the weighted $K_s$-band average
magnitude vs. the weighted standard deviation of the $K_s$ light curves, 
for 3 tiles at Galactic longitude $\ell =-2^{\rm o}$ and at different latitudes, 
namely $b = 0$ ({\em top}), $-5.2^{\rm o}$ ({\em middle}), and $-9^{\rm o}$ ({\em bottom}).
The plots illustrate the large changes in the
noise distribution of the light curves as an effect of increasing
source crowding towards the Galactic plane. The deviations from the
main locus at magnitudes $<12$~mag are due to (temporal) saturation. Note
that the color scale, mapped on the right, indicates the logarithm of the number 
of sources in each bin.}
\label{fig:hess}
\end{figure*}

The very complete VVV light curves for many different types of variable 
stars will also be useful as templates~~-- and indeed, many such reliably 
classified light curves with complete phase coverage have already been 
incorporated into the so-called {\em VVV Templates Database} 
\citep{raea14},\footnote{\tt http://vvvtemplates.org/} 
whose main purpose is to feed the classification algorithms that are 
currently being devised, in order to perform automated classification 
of the expected $10^{6-7}$ variable stars detected in the course of the 
VVV Survey \citep{mcea11,mcea13}. 
It is worth noting, in this regard, that our team has also 
recently completed a variable star catalog \citep{ceflea14} using data 
from the 3.8m United Kingdom Infrared Telescope's (UKIRT) Wide-Field near-IR 
CAMera (WFCAM; \citealt{mcea07}) calibration database. This multi-band 
database has also yielded high-quality lightcurve templates, which have 
also been incorporated into the VVV Templates Database \citep{raea14}.

\section{VVV-VSC: The VVV Variable Star Catalog}\label{sec:vvv-vsc}

VVV is a public survey, and so one of our main objectives is to provide 
a variable star catalog (VSC) that will ultimately be useful to the whole 
astronomical community. Since we last described our work on stellar 
variability in the framework of the VVV Survey \citep{mcea11,mcea13}, 
we have made much progress at building a comprehensive, science-ready 
relational database of variable VVV point sources. In its current 
incarnation, the VVV-VSC uses the high-quality aperture photometry 
provided by the Cambridge Astronomy Survey Unit (CASU; see \citealt{mcea13}, 
for more details and extensive references). The pipeline first extracts a 
master source list from the CASU tile catalogs, and then cross-matches 
those sources with the individual ``pawprints'' that are used to build 
each tile (six pawprints per tile; \citealt{dmea10}). 
Zero-point errors are then added, a threshold rejection procedure applied, 
aperture optimization carried out (by means of a minimization of the 
rms scatter), and descriptive statistics computed. Using a number of 
variability indices \citep[e.g.,][]{ws93,pbs96,ceflea14}, a list of variable 
star candidates is then finally obtained. Typically, we have been encountering
of order 50,000 candidates per VVV tile.

\begin{figure*}[!t]
\centering \includegraphics[width=\textwidth]{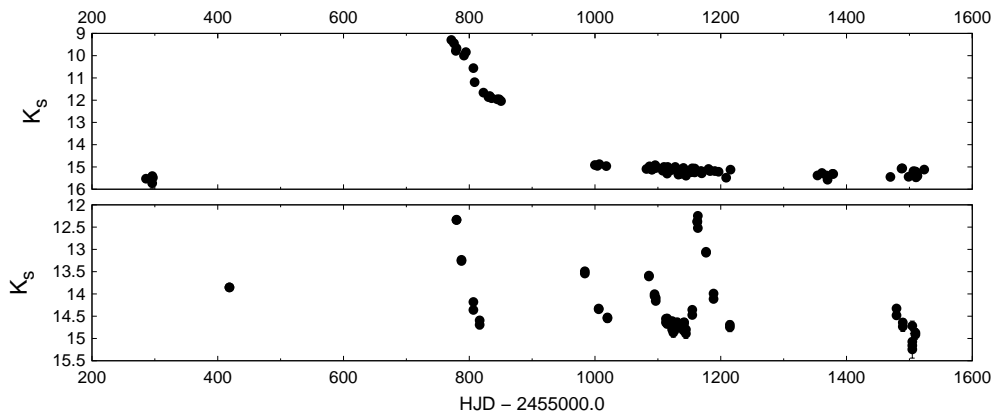}
\vskip0pt
\caption{Examples of transient sources in the VVV data. 
The {\em upper panel} shows the $K_s$-band light curve of VVV-WIT-02 \citep{idea14}, 
which is probably a recurrent nova. 
The {\em bottom panel} shows the VVV $K_s$-band light curve of the candidate
young stellar object (YSO) called SSTGLMC G353.4039-00.0664 \citep{trea08}, which is 
located in tile b329 (in the bulge plane), previously known from GLIMPSE. The
star has a large-amplitude recurring outburst with an approximate period of 
about 95~days.}
\label{fig:trans}
\end{figure*}

Current tests indicate that our procedure is highly successful at recovering
variable stars in general, and particularly periodic sources with sufficiently 
high amplitudes in the near-IR. At present, the procedure does not perform 
well for variable sources such as the following: very fast transients coming
from below detection; high-proper motion objects; extremely faint sources; 
very blue, faint sources; highly crowded sources; and very bright variables 
(i.e., $\langle K_s \rangle \lesssim 12$~mag), which are commonly saturated 
in the VVV images. Figure~\ref{fig:hess} illustrates the mean scatter of the 
recovered variable star candidates as a function of the mean $K_s$-band 
magnitude at three different Galactic latitudes, where the effects of 
saturation, crowding, and image depth are all apparent.

\begin{figure*}[!t]
\centering \includegraphics[width=\textwidth]{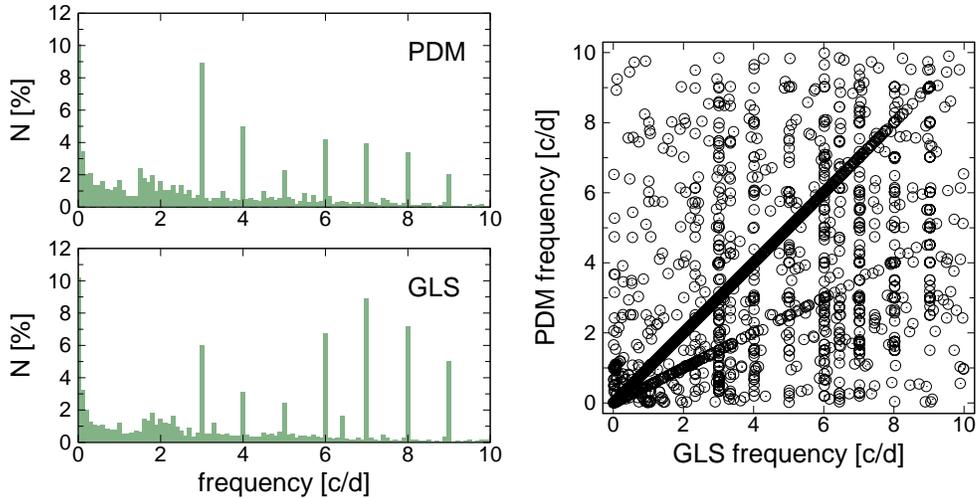}
\vskip0pt
\caption{({\em left panel}) 
This plot shows the histograms of the principal frequencies detected by the
two methods employed by the VVV variability pipeline, namely PDM ({\em top})
and GLS ({\em bottom}), after all the contributions 
from signals in the $10/T_{\rm tot}$ vicinity of integer c/d frequencies have
been removed~~-- where $T_{\rm tot}$ is the total time span covered by the data. 
The distributions are rather flat, but a small bump due to RR~Lyrae stars
can be spotted. The peaks at higher frequencies, close to
integers and half-integers, are spurious signals, caused mostly 
by the rotating diffraction spikes around bright stars. 
The {\em right panel} shows the consistency between the frequencies detected
by the two methods. Most are in agreement, as indicated by the high 
concentration of points around the 45-degree line 
overplotted on the data. There is, however, an additional sequence, with 
GLS periods that are half the PDM periods. This sequence is primarily due to 
eclipsing binary stars where the GLS algorithm recovers half the true orbital 
period, due to its diminished sensitivity to the presence of alternating peaks, 
compared to PDM.}
\label{fig:freqs}
\end{figure*}

Naturally, not all variables detected in the VVV data 
are periodic, and many different examples of variable stars that are 
not strictly periodic are shown, for instance, in 
\citet{mcea13} and \citet{rsea13}. However, many of the detected variables, such as 
Cepheids, RR~Lyrae, and eclipsing binary stars, {\em are} periodic~~-- and  
thus, once the variable stars are selected following the described procedure, 
we also carry out a frequency 
analysis, in order to find periodic and long-term signals, as may be present 
for each individual star. 
Figure~\ref{fig:trans} shows an example of non-periodic transient events 
detected in our data. 
To derive the periods, we currently use two different algorithms, 
namely the generalized Lomb-Scargle periodogram \citep[GLS;][]{zk09} and 
phase-dispersion minimization \citep[PDM;][]{rs78}. 

Figure~\ref{fig:freqs} compares the principal frequencies detected by the 
GLS and PDM methods. Most of the derived periods are in excellent agreement, 
as indicated by the large concentration of points around the 45-degree line 
that is overplotted in this diagram. However, a secondary sequence is also
present, representing favored GLS periods which are half 
as long as the corresponding PDM periods. Close inspection of the light 
curves of stars falling along this sequence reveals that it is mostly caused 
by eclipsing binary stars for which GLS incorrectly favors a period that is 
half as long as the true orbital period. This is caused by GLS's reduced sensitivity 
to the presence of alternating peaks in the time series, compared to PDM.  
Note also the presence of a large number of spurious detections at integer 
and half-integer periods. Most of these variable star candidates are actually 
{\em not} intrinsically variable, 
but represent instead stars in the vicinity of bright 
sources, whose spurious, periodic ``variability signal'' is actually produced 
by the rotating diffraction spike pattern that is seen around these very 
bright stars.  

By the end of the survey, once the VVV-VSC catalog is completed, we will be in a 
position to provide a relational database containing variability information 
for up to several million point sources. Our variability tables will include 
detailed information on the degree of variability, including for instance 
mean, median, minimum and maximum magnitudes, standard deviations, rms 
scatter, skewness, kurtosis, etc.; frequency analysis tables, including 
information on the power spectra and the periods derived therefrom; and 
queryable light curves and plots, enabling the display and retrieval of 
data (both raw and processed) in all 5 bandpasses used in the survey
(Sect.~\ref{sec:intro}).

\section{Conclusions}\label{sec:concl}

\noindent The VVV ESO Public Survey provides a treasure trove of scientific data that can be exploited in numerous different scientific contexts. In terms of stellar variability, the project will provide up to several million calibrated $K_s$-band light curves for genuinely variable sources, including pulsating stars, eclipsing systems, rotating variables, cataclysmic stars, microlenses, planetary transits, and even transient events of unknown nature. In this paper, we have provided a brief overview of the current status of the survey, as far as detection, cataloging, and classification of variable point sources is concerned. 

VVV is a public survey, and so the data will quickly be made available to the entire astronomical community as we move along, thus opening the door to many additional applications and synergies with other ongoing and future projects that target the same fields as those covered by VVV. In this sense, in this paper we have also described our current efforts at creating the VVV Variable Star Catalog (VVV-VSC)~~-- a fully queryable relational database that will allow users to quickly and efficiently retrieve variability and frequency analysis tables for the sources detected, multi-band color information, and~~-- last but not least~~-- the light curves themselves.


\vskip 1.0cm

\noindent {\bf Acknowledgements.} The VVV Survey is supported by the European Southern Observatory, 
the Basal Center for Astrophysics and Associated Technologies (PFB-06), and the Chilean Ministry 
for the Economy, Development, and Tourism's Programa Iniciativa Cient\'{i}fica Milenio through 
grant P07-021-F, awarded to The Milky Way Millennium Nucleus. 
M.C. gratefully acknowledges the support provided by Fondecyt through grant \#1141141.

\bibliographystyle{baaa}

\end{document}